\documentclass[reprint,
 amsmath,amssymb,
 aps,
prstab
]{revtex4-1}

\usepackage{graphicx}% Include Figure files
\usepackage{subcaption}
\usepackage{dcolumn}% Align table columns on decimal point
\usepackage{bm}% bold math
\usepackage{xcolor}
\usepackage{natbib} 
\usepackage[nottoc,numbib]{tocbibind}
\usepackage{amsmath}
\usepackage{lipsum}

\usepackage[%
  colorlinks=true,
  urlcolor=blue,
  linkcolor=blue,
  citecolor=blue
]{hyperref}
\usepackage{stackengine}
\usepackage{overpic}% super impose on the picture
\usepackage{pict2e}
\usepackage{tikz}

\usepackage{amsmath}

\usepackage{tabularx}
\usepackage{xcolor}

\usepackage{textcomp} %Textmu

\makeatletter
\newcommand\thefontsize[1]{{#1 The current font size is: \f@size pt\par}}
\newcommand\thefontsizeHere{{The current font size is: \f@size pt\par}}
\makeatother

\usepackage{tikz,xcolor,hyperref}

\definecolor{lime}{HTML}{A6CE39}
\DeclareRobustCommand{\orcidicon}{%
	\begin{tikzpicture}
	\draw[lime, fill=lime] (0,0) 
	circle [radius=0.16] 
	node[white] {{\fontfamily{qag}\selectfont \tiny ID}};
	\draw[white, fill=white] (-0.0625,0.095) 
	circle [radius=0.007];
	\end{tikzpicture}
	\hspace{-2mm}
}

\foreach \x in {A, ..., Z}{%
	\expandafter\xdef\csname orcid\x\endcsname{\noexpand\href{https://orcid.org/\csname orcidauthor\x\endcsname}{\noexpand\orcidicon}}
}

\begin{document}

\preprint{APS/123-QED}

\title{Filling Pattern Dependence of Regenerative Beam Breakup Instability in Energy Recovery Linacs}% Force line breaks with \\
%\title{An investigation of RF instabilities during beam loading of recirculating Energy Recovery Linac}% Force line breaks with \\
% Author Orchid ID: enter ID or remove command
\newcommand{\orcidauthorA}{0000-0002-5903-8930} % Sadiq; Add \orcidA{} behind the author's name
\newcommand{\orcidauthorB}{0000-0001-6346-5989} % Rob; Add \orcidA{} behind the author's name
\newcommand{\orcidauthorC}{0000-0002-8987-4999} % Peter; Add \orcidA{} behind the author's name
%https://orcid.org/0000-0002-8987-4999 % Peter

\author{S.~Setiniyaz\orcidA{}}
\email{s.saitiniyazi@lancaster.ac.uk} 

\author{R.~Apsimon\orcidB{}}
\email{r.apsimon@lancaster.ac.uk}
\affiliation{Engineering Department, Lancaster University, Lancaster, LA1 4YW, UK }
\affiliation{Cockcroft Institute, Daresbury Laboratory, Warrington, WA4 4AD, UK}

\author{P.~H.~Williams\orcidC{}}
\email{peter.williams@stfc.ac.uk}
\affiliation{STFC Daresbury Laboratory \& Cockcroft Institute, Warrington, WA4 4AD, UK}

\date{\today}% It is always \today, today,

\begin{abstract}
Beam breakup instability is a potential issue for all particle accelerators and is often the limiting factor for the maximum beam current that can be achieved. This is particularly relevant for Energy Recovery Linacs with multiple passes where a relatively small amount of charge can result in a large beam current. Recent studies have shown that the choice of filling pattern and recirculation scheme for a multi-pass energy recovery linac can drastically affect the interactions between the beam and RF system. In this paper we further explore this topic to study how filling patterns affect the beam breakup instability and how this can allow us to optimise the design in order to minimise this effect. We present a theoretical model of the beam-RF interaction as well as numerical modeling and show that the threshold current can vary by factors of 2-4, and potentially even more depending on the machine design parameters. Therefore a judicious choice of filling pattern can greatly increase the onset of BBU, expanding the utility of future ERLs.

\end{abstract}

\maketitle

%\begin{itemize}
%	\item short intro into recirculating ERLs
%	\subitem citing design studies around the world
%	\item filling patterns and BL patterns
%	\subitem number of filling patterns for an n-turn ERL
%	\subitem encoding of filling patterns and beam loading patterns
%	\subitem explain how some filling patterns are better or worse than others
%	\item beam loading simulation
%	\subitem explain dynamic and static set-points and show which one is better by simulation
%	\subitem show simulation results for different $S/N$ ratio
%	\subitem brute force method to identify optimal patterns
%	\subitem explain that the different set-points have different Figure of merit for optimal patterns 
%	\item optimal filling and BL patterns
%	\subitem explain that brute force isn't improved with extra computing power
%	\subitem study properties and characteristics of optimal (or potentially optimal) patterns
%	\subitem describe method for finding optimal patterns
%	\item results
%	\subitem use simulations to complement and verify studies
%	\subitem RF stability for odd and even pass ERLs
%	\subitem key implications of these results
%	\item acknowledgments
%\end{itemize}

\section{Introduction}
In the 2020 European Strategy for Particle Physics~\cite{EuropeanStrategyGroup:2020pow}, superconducting Energy Recovery Linacs (ERLs)~\cite{Merminga2016} were identified as a key accelerator technology requiring priority research and development to underpin the future anticipated needs of the community. Applications in particle physics cover both ERL-based colliders, such as LHeC~\cite{AbelleiraFernandez2012,LHeCnew}, PERLE~\cite{PERLE}, FCC-ee~\cite{Litvinenko:2019txu} and beam coolers for hadron colliders, notably EIC~\cite{Accardi2016}. ERLs are also seen as a promising option, in both academic and industrial contexts, for future Free-Electron laser light sources~\cite{Socol2013,Socol2011} and for nuclear physics through both direct beam internal target experiments~\cite{Hug:2017ypc} and through secondary production of narrowband gammas via inverse Compton scattering~\cite{Shimada2010,Hayakawa2010}. \par
One common theme in these future applications is the requirement of high average beam power, with GW being an aspired-for reasonable mid-term goal. This is three orders of magnitude beyond that achieved to date, namely by the JLab FEL upgrade~\cite{Neil:2000zz,Alarcon:2013yxa}, indicating the importance of addressing this as a priority for ERL research.\par
A well known limitation on the current one can support in an ERL is a particular incarnation of so-called beam breakup (BBU) instabilities. BBU has potential to occur where the beam interacts with a higher order mode (HOM) of an RF structure that it traverses and becomes deflected. There are two general classes of BBU, cumulative~\cite{Wangler2008, Chao2013, Padamsee1998, Gluckstern:1985yh, Delayen2003, Delayen2005} and regenerative~\cite{LYNEIS1983269, Pozdeyev2005}. Cumulative is where the deflection builds up over multiple structures and is not the subject of this study. Regenerative BBU is that of primary concern in recirculating and energy recovery linacs where there are multiple passes of the same bunch through each RF structure. This constitutes a feedback loop between the beam position offset and the HOM voltage.\par
Bunches passing through the cavity will excite a transverse HOM whose amplitude is dependent on the transverse offset of the passing bunches. In turn, the HOM voltage will deflect the subsequent bunches and increase their offsets. If the beam current is at a certain threshold, $I_{th}$, then the magnitude of the HOM voltage and bunch offsets will reach an equilibrium, below this the HOM voltage will tend to zero, above it will cause the HOM voltage and bunch offsets will grow exponentially until beam loss occurs. This defines the BBU instability~\cite{Altenmueller1966,Neil1970}. Whilst studies have been undertaken to investigate BBU instabilities for ERLs~\cite{Hoffstaetter2004,Hoffstaetter2019,Pozdeyev2006,Volkov2018,Tennant2005,TennantDissertation2006,Merminga2002}, no studies have yet considered the impact of different beam filling patterns and beam line topologies on the beam loading transients imparted on the cavities and bunches. This becomes important when considering multi-pass ERLs which, having now been demonstrated experimentally~\cite{Bartnik:2020pos}, feature in many designs for future ERL facilities due to the obvious advantage of providing higher beam energies without concomitant increase in the number of accelerating structures. The effect of the filling pattern, which describes the order in which bunches are injected into the ERL over subsequent turns, has in a previous publication~\cite{Setiniyaz2020} been shown to have a major impact on the stability of the RF system. Just as the fundamental mode and RF power are affected differently by our choice of beam line topology and filling pattern, the threshold current of regenerative BBU is also dependent on these parameters. In the following section, we show that the impact on BBU is somewhat more complex than it is for the fundamental mode due to the asynchronous nature of the HOM mode relative to the beam, as well as the transcendental relationship between the HOM voltage and the bunch offsets.\par
For regenerative BBU, the HOM mode has a sufficiently high Q-factor that the mode persists for a relatively long timescale. As the bunches pass through the structure they are given a transverse kick, whilst also contributing to the excitation of the HOM due to their off-axis trajectories. On recirculation they pass back through the structure with a larger offset, further exciting the mode. This feedback loop can grow exponentially until beam loss occurs.\par
\begin{figure}
\scalebox{0.45} [0.45]{\includegraphics{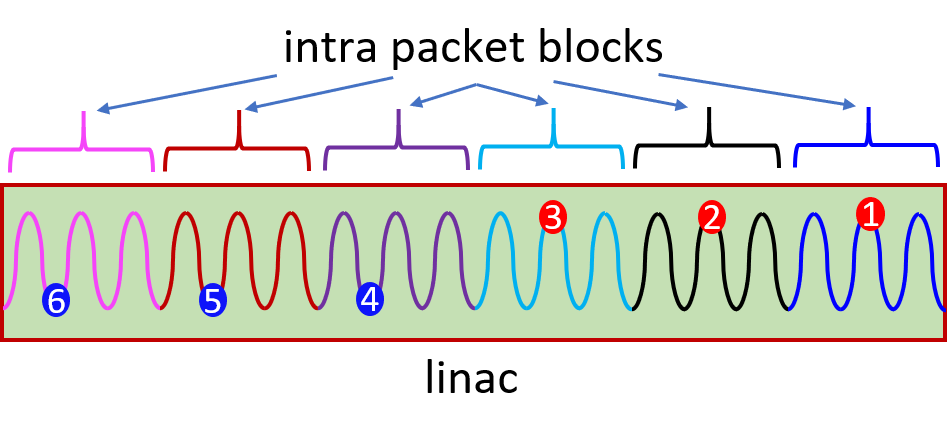}}
	\caption{Intra-packet blocks. RF cycles are shown within the linac. Each block is colored differently. The red/blue bunches are on the peak/trough and being accelerated/decelerated.}
	\label{fig:Blocks} 
\end{figure}
\begin{figure}
\scalebox{0.45} [0.45]{\includegraphics{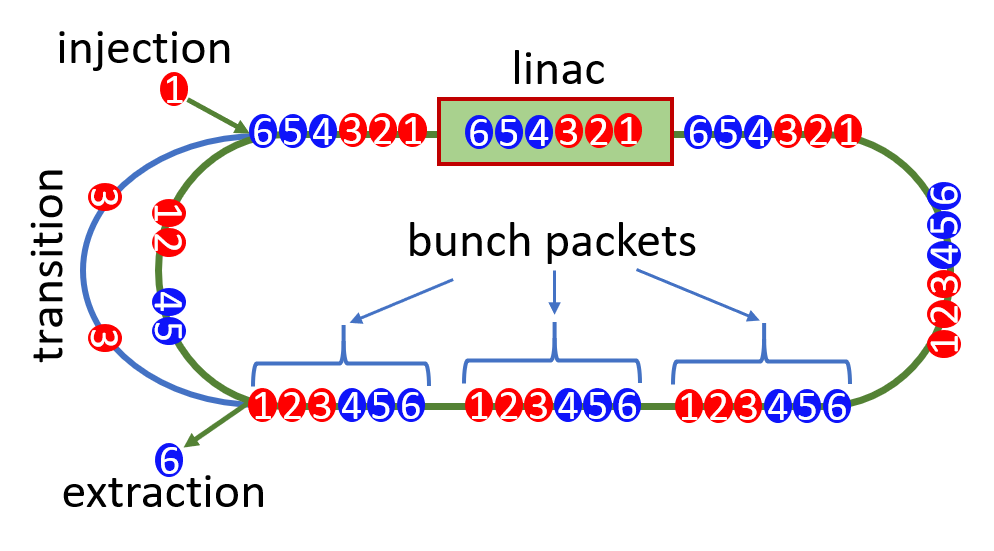}}
	\caption{Bunch packets in the ERL. Bunches at the third turn go through a transition arc where there is at least an extra half RF cycle delay. }
	\label{fig:PacketsInERL} 
\end{figure}

%For cumulative BBU, the HOM mode is typically short-lived, but has a high $R/Q$, which allows for a strong coupling between the HOM and the beam. As such, when the bunch passes through the structure, it receives a strong transverse kick due to the HOM and high $R/Q$, which can build up over many structures. However regenerative BBU is often the main concern for recirculating machines and will be the focus of this paper.
Filling patterns are used to describe the order in which bunches are injected into a ring on subsequent turns, here we use the concept (introduced in~\cite{Setiniyaz2020}) of intra-packet blocks to describe the position they occupy on each turn. For example, for a 6-turn ERL (3 accelerating passes and 3 decelerating), the filling pattern [1~2~3~4~5~6] indicates that the first bunch goes to the first intra-packet block, the second bunch goes to the second block, and so on. The intra-packet block also fixes how many RF cycles are between bunches, as in the general case bunches are not necessarily injected into every RF cycle, illustrated in Fig.~\ref{fig:Blocks}. These bunches forms a packet and multiples of such packets fill up the ring as shown in the Fig.~\ref{fig:PacketsInERL}. Similarly, filling pattern [1~3~2~4~5~6] indicates that the first bunch goes to the first block, second bunch goes to the third block, and so on. These filling schemes are called FIFO (first in first out) as the bunches maintain their order in the packet, however, their turn number changes turn by turn. One can also generated a packet where turn number in the packet doesn't change turn by turn by and consequently the cavity always see same packets passing thorough. Such a scheme is called sequence preserving (SP) scheme and denoted with curly brackets to be differentiated from square brackets of FIFO schemes. For example, SP pattern $\{1~4~2~5~3~6\}$ describes the first bucket is always occupied by bunch at the first turn, the second bucket is always occupied by the bunch at fourth turn, and so on. The key differences are that in FIFO schemes, bunches remain in the same RF block and bunch turn numbers change, whilst in SP schemes the arc length for each turn of the ERL is specifically designed such that the bunches transition between RF blocks on each turn such that the sequence of bunch turn number in a packet remains constant on each turn; hence the name sequence preserving.

The number of possible filling patterns increases as $\left(N-1\right)!$, where $N$ is the number of recirculation passes of the ERL. The filling patterns are dependent on the bunch injection and recombination schemes. FIFO and SP patterns are merely the most simple subsets of all possible schemes that can be generated, this will be further studied in a future work employing a group theoretic classification technique. In this work we focus on SP schemes to show that BBU is affected by the filling pattern. There are several factors in multi-turn ERLs that can affect BBU, which are not present in circular machines: bunches at different turns have different energies so under the same HOM voltage higher energy bunches are deflected less than low energy ones; bunches transitioning from accelerating phase to decelerating phase need to be delayed by at least half an RF cycle and thus have a changed revolution time; bunches accumulate offset and kicks within the recirculating process which become amplified as the bunch energy  decreases, this is termed adiabatic anti-damping. All these effects must be taken into account when determining filling pattern dependent BBU threshold currents. 

In section II, we present an analytical derivation for the threshold current and its dependence on filling patterns and beam line topology and consider how it drives BBU. Due to the transcendental relationship between HOM voltage and beam trajectories, the analytical approach is unable to provide a simple calculation of the threshold current, however we are able to exploit it to provide insight into the behaviour of BBU under certain conditions. In Section III, we provide the results of numerical simulations from Matlab in order to determine the threshold current for different filling patterns for the SP schemes. These numerical simulations allow us to look at the frequency dependence of the threshold current for different filling patterns as well as to determine the threshold current for all filling patterns as a frequency-averaged threshold current. From this, we are able to show that with the correct choice of filling pattern, the threshold current can be increased significantly with appropriate optimisation.

\section{Analytical Model}
In this section, we derive an analytical expression for the threshold current to allow us to understand how the filling pattern and recirculation scheme of the ERL affects the threshold current. We start by following the standard derivation described in other works, such as~\cite{Pozdeyev2005,TennantDissertation2006,Pozdeyev2006}. Then we augment this initial expression for the threshold current, casting it in terms of a generic beam filling pattern.\par
BBU can be a long and/or short-range effect and is caused by transverse offsets of the bunch centroids exciting deflecting modes, which are usually dipole fields, which then give a transverse kick to the bunch on the next turn. For short-range BBU, the HOM mode excited has a small loaded Q-factor ($Q_{L}$) and as such the mode is sufficiently attenuated before the bunches return in order to prevent a significant transverse kick to the bunch after it has completed one revolution, but it can have a significant effect on bunches directly behind the driving bunch. From the Panofsky-Wenzel theorem~\cite{pw}, we relate the transverse voltage of the HOM mode in terms of the transverse variation in longitudinal voltage as:

\begin{equation}
    V_{\perp}=-\frac{ic}{\omega}\frac{V_{z}\left(x\right)}{x}e^{i\phi}
\label{eq:Vtrans}.
\end{equation}

It can also be shown that a bunch of charge $q_{bunch}$ excites this HOM mode and increases the transverse voltage as

\begin{equation}
    \delta{V}_{\perp}=-i\frac{q_{bunch}\omega}{2}\left(\frac{\omega{x}}{c}\right)\left(\frac{R}{Q}\right)_{\perp}e^{i\phi}
\label{eq:Vtrans1}
\end{equation}

\noindent where $x$ is the transverse offset from the electrical centre of the cavity, $c$ is the speed of light, $\omega$ is the angular frequency of the HOM mode, $\left(\frac{R}{Q}\right)_{\perp}$ is the transverse $R/Q$ and $\phi$ is the phase of the beam with respect to the peak of the HOM voltage. A derivation of this beam loading can be found in \cite{apsimon}.

For a recirculating ERL, the energy of the bunches will vary from one turn to the next and therefore, we need to consider the one-turn map for each turn, taking into account the energy variation of the bunch during the turn. Furthermore, we also need to factor in the deflection from the HOM on each turn, which will depend on the bunch energy. In general, we can write a simple first-order one-turn map as

\begin{equation}
\begin{array}{l}
    \begin{pmatrix}
    x_{n} \\ x'_{n}
    \end{pmatrix}=\begin{pmatrix} R_{11}^{\left(n\right)} & R_{12}^{\left(n\right)} \\ R_{21}^{\left(n\right)} & R_{22}^{\left(n\right)}\end{pmatrix}\begin{pmatrix}
    x_{n-1} \\ x'_{n-1}
    \end{pmatrix}+ \begin{pmatrix}
    \delta{x}_{HOM}^{\left(n\right)} \\ \delta{x'}_{HOM}^{\left(n\right)}
    \end{pmatrix}
    \end{array}
    \label{eq:turnmap}
\end{equation}

\noindent where $R_{ij}^{\left(n\right)}$ is the $i,j$ matrix element for the one-turn map for turn $n$, $x_{n}$ and $x'_{n}$ are the position and angle trajectory of the bunch on turn $n$ and $\delta{x}_{HOM}^{\left(n\right)}$ and $\delta{x'}_{HOM}^{\left(n\right)}$ are the position and angle deflection from the HOM mode. From Eq.~\ref{eq:turnmap}, we have a means of determining all the bunch positions if the transverse HOM voltage is known. In addition, from Eq.~\ref{eq:Vtrans1}, we can write an expression for the collective beam loading of the HOM mode due to $k$ bunches, taking into account the $Q_{L}$ of the mode and the phase shift between bunches as

\begin{equation}
    \delta{V}_{\perp,k}=-i\frac{\omega^{2}}{2c}\left(\frac{R}{Q}\right)_{\perp}e^{\left(-\frac{\omega{}T_{k+1}}{2Q_{L}}+i\phi_{0}\right)}\sum_{j=1}^{k}{q_{j}x_{j}e^{\omega{T_{j}}\left(\frac{1}{2Q_{L}}+i\right)}}
\label{eq:Vtrans2}.
\end{equation}

In Eq.~\ref{eq:Vtrans2}, we define without loss of generality that $T_{k+1}=0$, hence $e^{-\frac{\omega{T}_{k+1}}{2Q_{L}}}=1$. The stored energy in a cavity is given as

\begin{equation}
    U=\frac{\left|V_{\perp}\right|^{2}}{\omega\left(\frac{R}{Q}\right)_{\perp}}
\label{eq:Vtrans3},
\end{equation}

\noindent from Eq.~\ref{eq:Vtrans3}, we can determine the RF power transferred by the beam as

\begin{equation}
    \frac{\delta{U}}{\delta{t}}=\frac{2\left|V_{\perp}\right|}{\omega\left(\frac{R}{Q}\right)_{\perp}}\frac{\left|\delta{V}_{\perp}\right|}{\delta{t}}
\label{eq:Vtrans4}.
\end{equation}

By substituting Eq.~\ref{eq:Vtrans2} into Eq.~\ref{eq:Vtrans4} and assuming that the bunch charge is either constant for all bunches or periodic such that the mean bunch charge is bounded, then $\sum_{j=1}^{k\rightarrow\infty}{\frac{q_{j}}{\delta{t}}}=I_{beam}$, then we obtain:

\begin{equation}
    \frac{dU_{beam}}{dt}=I_{beam}\frac{\omega\left|V_{\perp}\right|}{c}\left|\sum_{j=1}^{k}{x_{j}e^{\omega{T_{j}}\left(\frac{1}{2Q_{L}}+i\right)}}\right|
\label{eq:Vtrans5}.
\end{equation}

We also know that when the cavity HOM voltage is at equilibrium,

\begin{equation}
    \frac{dU_{cav}}{dt}=\frac{dU_{beam}}{dt}+P_{c}=0
\label{eq:Vtrans6},
\end{equation}

\noindent where $P_{c}$ is the ohmic power dissipation in the cavity, given as

\begin{equation}
    P_{c}=\frac{\left|V_{\perp,k-1}\right|^{2}}{2\left(\frac{R}{Q}\right)_{\perp}Q_{L}}
\label{eq:Vtrans7}.
\end{equation}

The threshold current is defined as the maximum allowed beam current before the HOM voltage grows exponentially, thus it is the limit of stability for the accelerator. At the threshold current, the HOM voltage is at steady state at some value, below it, the HOM voltage will decay exponentially. Thus the only solutions to Eq.~\ref{eq:Vtrans6} are the trivial case where $\left|V_{\perp}\right|=0$, or when $I_{beam}=I_{th}$, where $I_{th}$ is the threshold current. We can obtain the threshold current from Eqs.~\ref{eq:Vtrans6} and \ref{eq:Vtrans7} as

\begin{equation}
    I_{th}=I_{beam}=-\frac{\left|V_{\perp}\right|c}{2\left(\frac{R}{Q}\right)_{\perp}\omega{Q}_{L}\left|\sum_{j=1}^{k}{x_{j}e^{\omega{T_{j}}\left(\frac{1}{2Q_{L}}+i\right)}}\right|}
\label{eq:Vtrans8}.
\end{equation}

From hereon, it is convenient to introduce some definitions as the beam loading will be periodic over several timescales. We define a bunch packet as an ensemble of $N$ consecutive bunches for an $N$-turn ERL, such that each bunch has completed a different number of recirculating passes. Furthermore, it is assumed that on a given turn, each bunch packet is equivalent. Finally, we know that from Eq.~\ref{eq:turnmap} that if the bunch offsets are dominated by the kicks from the HOM, then $x_{j}$ will depend on the HOM voltage, $V_\perp$, which in turn will depend on the bunch charge, $q_{j}$. If we assume that the variation in $q_{j}$ is small, then it can be approximated as a constant. Thus if $V_\perp$ and $x_{j}$ both depend on the bunch charge, then the overall charge dependence is cancelled in the threshold current. As we shall see further on in this section, the threshold current will have a charge in the equation, although as we know that the charge dependence cancels, we can take $q=1$~C for convenience without loss of generality.

From~\cite{Setiniyaz2020} it is known that for any choice of bunch filling pattern and recirculation scheme, the filling pattern will repeat every $N$ turns, or in some cases an integer divisor of $N$. Additionally, in~\cite{Setiniyaz2020} the concept of a sequence preserving (SP) scheme is discussed, whereby for a given filling pattern, a recirculation scheme can be chosen such that the sequence of bunch turn numbers in a packet is the same on each turn. For SP schemes, we see that the bunch packets are equivalent for all turns and so we can also trivially see that this will indeed be periodic over $N$ turns. To aid in the following derivation, we shall define a form factor for a single bunch packet ($f_{n,k}$), a single turn ($F_{n,k}$), and $N$-turns ($\langle{F_{k}\rangle}$) respectively as

\begin{equation}
\begin{array}{l}
    f_{n,k} = e^{-\frac{\omega{T}_{N+k}}{2Q_{L}}}\left|\sum_{j=k}^{N+k-1}{x_{j}e^{\omega{T}_{j}\left(\frac{1}{2Q_{L}}+i\right)}}\right| \\
    F_{n,k} = e^{-\frac{\omega{T}_{MN+k}}{2Q_{L}}}\left|\sum_{j=k}^{MN+k-1}{x_{j}e^{\omega{T}_{j}\left(\frac{1}{2Q_{L}}+i\right)}}\right| \\
    F_{k} = e^{-\frac{\omega{T}_{MN^{2}+k}}{2Q_{L}}}\left|\sum_{j=k}^{MN^{2}+k-1}{x_{j}e^{\omega{T}_{j}\left(\frac{1}{2Q_{L}}+i\right)}}\right|
\end{array}
\label{eq:Vtrans9},
\end{equation}

\noindent where $k$ is the bunch number in the packet, $n$ is the turn number. These form factors are a convenient means of quantifying the effect on HOM voltage from the bunches over different timescales. These form factors depend explicitly on the filling pattern and recirculation scheme.

As we assume that the bunch structure is periodic over a bunch packet and also periodic over $N$ turns, then it is also convenient to define the duration of a bunch packet as $T_{packet}$ and the mean revolution period of the ring as $T_{rev}=MT_{packet}$. Based on this, we can express the 1-turn and $N$-turn form factors in terms of the bunch packet form factor as

\begin{equation}
\begin{array}{l}
    f_{n,k} = e^{-\frac{\omega\left(T_{packet}+T_{k}\right)}{2Q_{L}}}\left|\sum_{j=k}^{N+k-1}{x_{j}e^{\omega{T}_{j}\left(\frac{1}{2Q_{L}}+i\right)}}\right| \\
    F_{n,k} = \left|\sum_{m=0}^{M-1}{e^{m\omega{T}_{packet}\left(\frac{1}{2Q_{L}}+i\right)}}\right|f_{n,k} \\
    F_{k} = \left|\sum_{m=0}^{M-1}{e^{m\omega{T}_{packet}\left(\frac{1}{2Q_{L}}+i\right)}}\right|\sum_{n=1}^{N}{f_{n,k}e^{\frac{\left(n-N\right)\omega{T}_{rev}}{2Q_{L}}}}
\end{array}
\label{eq:Vtrans10}.
\end{equation}

We note in Eq.~\ref{eq:Vtrans10} that the summations for the 1- and $N$-turn form factors are finite geometric sums which can easily be evaluated. From Eq.~\ref{eq:Vtrans8} it is convenient to convert the summation in the denominator into a more appropriate form and we shall do so by exploiting the $N$-turn form factor from Eq.~\ref{eq:Vtrans10} and will consider the summation in Eq.~\ref{eq:Vtrans8} to tend to infinity.

\begin{equation}
\begin{aligned}
    \left|\sum_{j=k}^{\infty}{x_{j}e^{\omega{T_{j}}\left(\frac{1}{2Q_{L}}+i\right)}}\right|= F_{k}\left|\sum_{n=0}^{\infty}{e^{-n\omega{T}_{rev}\left(\frac{1}{2Q_{L}}+i\right)}}\right| \\
    = \frac{F_{k}e^{\frac{N\omega{T}_{rev}}{4Q_{L}}}}{\sqrt{2}\sqrt{\cosh{\left(\frac{N\omega{T}_{rev}}{2Q_{L}}\right)}-\cos{\left(N\omega{T}_{rev}\right)}}}
\end{aligned}
\label{eq:Vtrans11}.
\end{equation}

Next we shall evaluate the HOM voltage from Eq.~\ref{eq:Vtrans8} and for this, we shall assume that the cavity is at steady state, thus over a single turn, the HOM voltage amplitude will be periodic over each bunch packet, however for convenience, we shall determine the HOM voltage from all the bunches in a single turn. Thus we can say that the condition for convergence is given as

\begin{equation}
\begin{array}{l}
    V_{\perp}= V_{\perp}e^{-\omega{T}_{rev}\left(\frac{1}{2Q_{L}}-i\right)}\\
    -i\frac{\omega^{2}}{2c}\left(\frac{R}{Q}\right)_{\perp}e^{-\frac{\omega{T}_{rev}}{2Q_{L}}}e^{i\phi_{0}}\sum_{j=k-1}^{MN+k-2}{q_{j}x_{j}e^{\omega{T}_{j}\left(\frac{1}{2Q_{L}}+i\right)}}
\end{array}
\label{eq:Vtrans12}.
\end{equation}

We can rearrange Eq.~\ref{eq:Vtrans12} and solve for $V_{\perp}$, but we also only need the magnitude of this. We shall once again assume that the bunch charge is constant, thus we can consider the mean bunch charge and we can also express the summation as the 1-turn form factor from Eq.~\ref{eq:Vtrans11} to obtain

\begin{equation}
    \left|V_{\perp}\right|= \frac{\langle{q}\rangle\frac{\omega^{2}}{2c}\left(\frac{R}{Q}\right)_{\perp}e^{-\frac{\omega{T}_{rev}}{4Q_{L}}}F_{n,k-1}}{\sqrt{2}\sqrt{\cosh{\left(\frac{\omega{T}_{rev}}{2Q_{L}}\right)}-\cos{\left(\omega{T}_{rev}\right)}}}
\label{eq:Vtrans13}.
\end{equation}

We can now substitute Eqs.~\ref{eq:Vtrans11} and \ref{eq:Vtrans13} into Eq.~\ref{eq:Vtrans8} to obtain

\begin{equation}
\begin{array}{l}
    I_{th}= \\
    -\frac{\langle{q}\rangle\omega}{4Q_{L}}\sqrt{\frac{\cosh{\left(\frac{N\omega{T}_{rev}}{2Q_{L}}\right)}-\cos{\left(N\omega{T}_{rev}\right)}}{\cosh{\left(\frac{\omega{T}_{rev}}{2Q_{L}}\right)}-\cos{\left(\omega{T}_{rev}\right)}}}\frac{F_{n,k-1}e^{-\frac{\left(N+1\right)\omega{T}_{rev}}{4Q_{L}}}}{F_{k}}
\end{array}
\label{eq:Vtrans14}.
\end{equation}

It should be noted that in general the 1-turn form factor changes turn by turn, therefore this implies that the threshold current is time-dependent as a result of this. We also have that the form factor in the numerator is counting from the $\left(k-1\right)^{\text{th}}$ bunch, whereas in the denominator it is counting from the $k^{\text{th}}$ bunch. We can express the 1- and $N$-turn form factors in terms of the bunch packet form factors and sum the threshold current over all bunches and turns to give the following general expression for the average threshold current

\begin{equation}
\begin{array}{l}
    \langle{I_{th}}\rangle= -\frac{\alpha}{N^{2}}\frac{\langle{q}\rangle\omega}{4Q_{L}}\sum_{k=1}^{N}{\frac{\sum_{n=1}^{N}{f_{n,k-1}}}{\sum_{n=1}^{N}{f_{n,k}e^{\frac{n\omega{T}_{rev}}{2Q_{L}}}}}} \\ \\
    \alpha = e^{\frac{\left(N-1\right)\omega{T}_{rev}}{4Q_{L}}}\sqrt{\frac{\cosh{\left(\frac{N\omega{T}_{rev}}{2Q_{L}}\right)}-\cos{\left(N\omega{T}_{rev}\right)}}{\cosh{\left(\frac{\omega{T}_{rev}}{2Q_{L}}\right)}-\cos{\left(\omega{T}_{rev}\right)}}}
\end{array}
\label{eq:Vtrans15}.
\end{equation}

In Eq.~\ref{eq:Vtrans15}, $\frac{\langle{q}\rangle\omega}{4Q_{L}}$ is essentially the most basic possible estimate for the threshold current, $\alpha$ is an enhancement factor that depends on the HOM frequency, mean revolution period and number of recirculation turns in the ERL.

\subsection{Worked example: SP schemes}
As previously mentioned and also discussed in more detail in~\cite{Setiniyaz2020}, SP schemes have the property that the sequence of bunch turn numbers in a packet does not change from one turn to the next. It can be shown that all filling patterns will have a unique recirculation scheme to form an SP scheme, but the converse is not necessarily true. We can use the properties of SP schemes to find a special case form of Eq.~\ref{eq:Vtrans15} because the filling pattern and therefore bunch packet form factors are independent of turn number.

\begin{equation}
\begin{array}{l}
    \langle{I_{th}}\rangle= -\frac{\alpha'}{N}\frac{\langle{q}\rangle\omega}{4Q_{L}}\sum_{k=1}^{N}{\frac{f_{k-1}}{f_{k}}} \\ \\
    \alpha' = \frac{\sinh{\left(\frac{\omega{T}_{rev}}{2Q_{L}}\right)}}{\sinh{\left(\frac{N\omega{T}_{rev}}{2Q_{L}}\right)}}\sqrt{\frac{\cosh{\left(\frac{N\omega{T}_{rev}}{2Q_{L}}\right)}-\cos{\left(N\omega{T}_{rev}\right)}}{\cosh{\left(\frac{\omega{T}_{rev}}{2Q_{L}}\right)}-\cos{\left(\omega{T}_{rev}\right)}}}
\end{array}
\label{eq:Vtrans16}.
\end{equation}

\begin{figure}
\scalebox{0.23} [0.23]{\includegraphics{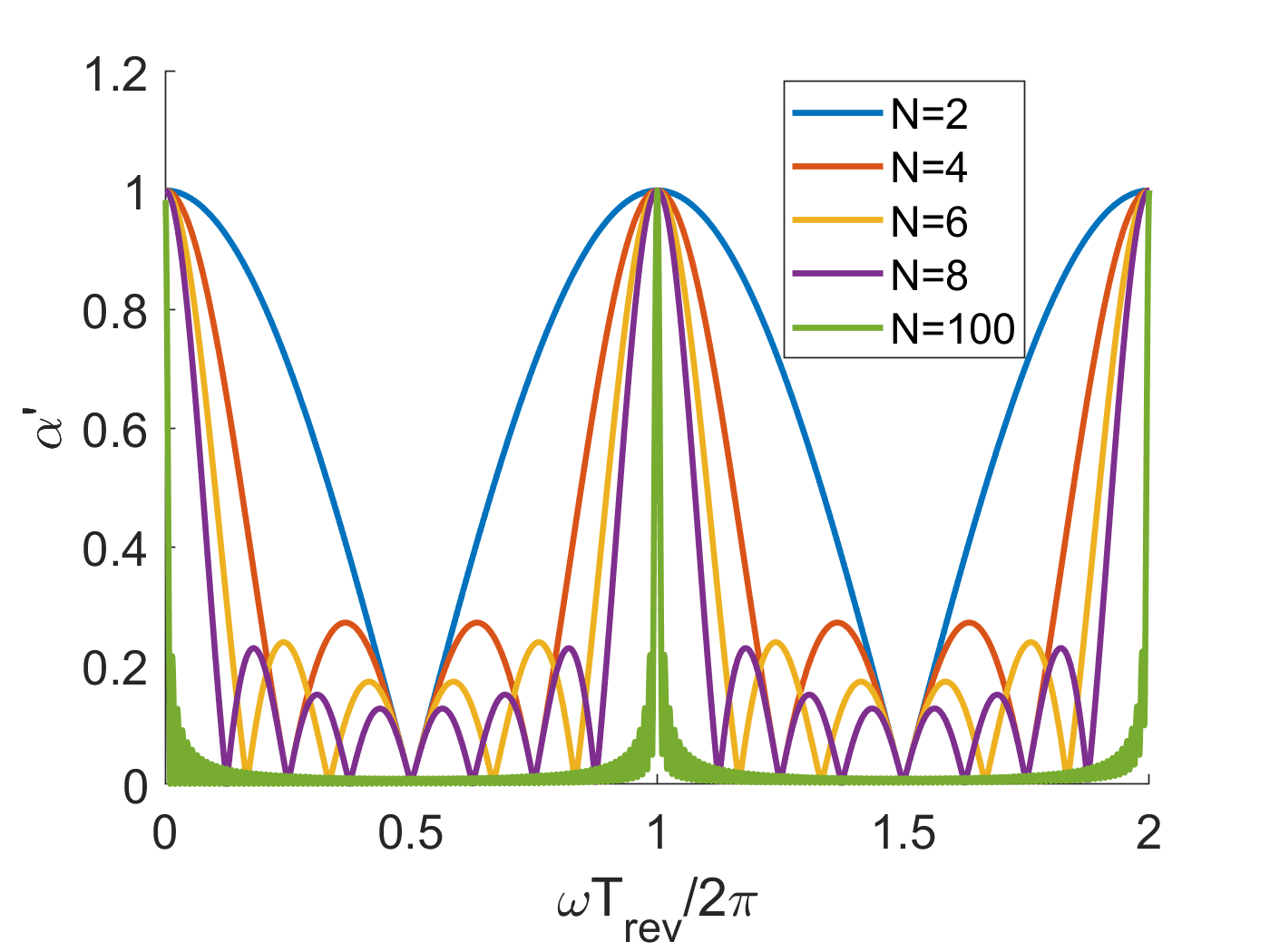}}
	\caption{$\alpha'$ vs. $\omega{T}_{rev}/2\pi$ for SP schemes for different numbers of recirculation turns, with $Q_{L}$ assumed to be 1000.}
	\label{Fig:BBUalpha}
\end{figure}

The ratio of form factors needs to be determined numerically due to the cyclic dependence between $\left|V_{\perp}\right|$ and the transverse offset of the beam (Eqs.~\ref{eq:turnmap} and ~\ref{eq:Vtrans13}), resulting in a transcendental equation for the cavity voltage. Fig.~\ref{Fig:BBUalpha} shows $\alpha'$ vs $\omega{T}_{rev}$ for different numbers of recirculation turns ($N$) where we have assumed that $Q_{L}=1000$. This shows that we expect the mean threshold current to decrease with increasing $N$, placing a constraint on the feasibility of designing arbitrarily many turn ERLs. Furthermore, it shows that $\alpha'$ tends to a series of Dirac $\delta$-functions in the limit that $N\rightarrow\infty$. As expected, $\alpha'$ is maximal when $\omega{T}_{rev}=2n\pi$, where $n$ is an integer, and also $\alpha'\approx0$ when $\omega{T}_{rev}=\left(2n+1\right)\pi$.

\subsection{Properties of the bunch packet form factor, $f_{n,k}$}
From Eqs.~\ref{eq:Vtrans15}, we note that the threshold current depends on both $f_{n,k}$ and $f_{n,k-1}$. This is because $\left|V_{\perp}\right|$ at some moment in time depends on the cumulative beam loading of all previous bunches; which depends on $f_{n,k-1}$. However, the threshold current also depends on the effect this voltage has on the next bunch passing through; which depends on $f_{n,k}$. It would therefore be beneficial to understand the relationship between the form factor evaluated over different bunches as well as to gain some insight into the structure of the form factor. Starting from Eq.~\ref{eq:Vtrans10}, it is useful to express the form factor in a non-modular form. Starting with the modular part, we can rewrite it as

\begin{equation}
\left|\sum_{j=k}^{N+k-1}{x_{j}e^{\omega{T}_{j}\left(\frac{1}{2Q_{L}}+i\right)}}\right|=\left|\sum_{j=k}^{N+k-1}{x_{j}e^{\frac{\omega{T}_{j}}{2Q_{L}}}e^{i\omega{T}_{j}}}\right|
\label{eq:Vtrans17}.
\end{equation}

\noindent We can now remove the modular form since $\left|z\right|^{2}=zz^*$, and in addition, we can write the multiple of two summations as a double-summation instead to obtain

\begin{equation}
\begin{array}{l}
\left|\sum_{j=k}^{N+k-1}{x_{j}e^{\omega{T}_{j}\left(\frac{1}{2Q_{L}}+i\right)}}\right|^{2}= \\
\sum_{j=k}^{N+k-1}{\sum_{l=k}^{N+k-1}{x_{j}x_{l}e^{\frac{\omega\left({T}_{j}+T_{l}\right)}{2Q_{L}}}e^{i\omega\left({T}_{j}-T_{l}\right)}}}
\end{array}
\label{eq:Vtrans18}.
\end{equation}

From Eq.~\ref{eq:Vtrans18}, we can split the double summation into two separate parts, the first where $j=l$ will become a single summation and for the case where $j\neq{l}$, due to the symmetry of the equation, when know that $f\left(j,l\right)=f\left(l,j\right)$, thus without loss of generality, we can take double the summation and impose the constraint that $j>l$ to obtain

\begin{equation}
\begin{array}{l}
\left|\sum_{j=k}^{N+k-1}{x_{j}e^{\omega{T}_{j}\left(\frac{1}{2Q_{L}}+i\right)}}\right|^2=
\sum_{j=k}^{N+k-1}{x_{j}^{2}e^{\frac{\omega{T}_{j}}{Q_{L}}}}+ \\
2\sum_{j=k+1}^{N+k-1}{\sum_{l=k}^{j-1}{x_{j}x_{l}e^{\frac{\omega\left({T}_{j}+T_{l}\right)}{2Q_{L}}}\cos{\omega\left({T}_{j}-T_{l}\right)}}}
\end{array}
\label{eq:Vtrans19}.
\end{equation}

Eq.~\ref{eq:Vtrans19} can be substituted back into Eq.~\ref{eq:Vtrans10} to give a convenient means of calculating the form factors without needing to compute complex numbers. Furthermore, in this form, we can explore the relationship between $f_{n,k}$ and $f_{n,k-1}$ by understanding how each summation over $k-1$ relates to the summation over $k$. We shall assume that the ERL is stable and periodic over a bunch packet, thus $x_{N+k}=x_{k}$ and also $T_{N+k}=T_{packet}+T_{k}$. From the single summation, we can express it as

\begin{equation}
\begin{array}{l}
\sum_{j=k-1}^{N+k-2}{x_{j}^{2}e^{\frac{\omega{T}_{j}}{Q_{L}}}}=\sum_{j=k}^{N+k-1}{x_{j}^{2}e^{\frac{\omega{T}_{j}}{Q_{L}}}}- \\ \\
2x_{k}^{2}e^{\frac{\omega{T}_{k}}{2Q_{L}}}e^{\frac{\omega{T}_{packet}}{2Q_{L}}}\sinh{\frac{\omega{T}_{packet}}{2Q_{L}}}
\end{array}
\label{eq:Vtrans20}.
\end{equation}

Applying the same to the double summation, we obtain

\begin{equation}
\begin{array}{l}
\sum_{j=k}^{N+k-2}{\sum_{l=k-1}^{j-1}{x_{j}x_{l}e^{\frac{\omega\left({T}_{j}+T_{l}\right)}{2Q_{L}}}\cos{\omega\left({T}_{j}-T_{l}\right)}}}= \\
\sum_{j=k+1}^{N+k-1}{\sum_{l=k}^{j-1}{x_{j}x_{l}e^{\frac{\omega\left({T}_{j}+T_{l}\right)}{2Q_{L}}}\cos{\omega\left({T}_{j}-T_{l}\right)}}} - \\ \\
r\sum_{j=k}^{N+k-2}{x_{j}x_{k}e^{\frac{\omega\left(T_{j}+T_{k}\right)}{2Q_{L}}}}\cos{\left(\omega\left(T_{k}-T_{j}\right)+\psi\right)}
\end{array}
\label{eq:Vtrans21}.
\end{equation}

\noindent where
\begin{equation}
\begin{array}{l}
r=\sqrt{2}e^{\frac{\omega{T}_{packet}}{4Q_{L}}}\sqrt{\cosh{\left(\frac{\omega{T}_{packet}}{2Q_{L}}\right)}-\cos{\omega{T}_{packet}}} \\
\text{and} \\
\psi=\tan^{-1}{\left(\frac{\sin{\omega{T}_{packet}}}{\cos{\omega{T}_{packet}}-e^{-\frac{\omega{T}_{packet}}{2Q_{L}}}}\right)}
\end{array}
\label{eq:Vtrans22}.
\end{equation}

Substituting Eqs.~\ref{eq:Vtrans19}-\ref{eq:Vtrans22} into Eq.~\ref{eq:Vtrans10}, we obtain

\noindent where

\begin{equation}
\begin{array}{l}
f_{n,k-1}^{2}= e^{\frac{\omega\left(T_{k}-T_{k-1}\right)}{Q_{L}}}\left(f_{n,k}^{2}\right. \\ \\
\left.-2x_{k}^{2}e^{-\frac{3\omega{T}_{packet}}{2Q_{L}}}\sinh{\left(\frac{\omega{T}_{packet}}{2Q_{L}}\right)}-2rx_{k}e^{-\frac{\omega{T}_{packet}}{Q_{L}}}\times\right. \\ \\
\left.\sum_{j=k}^{N+k-2}{x_{j}e^{\frac{\omega\left(T_{j}+T_{k}\right)}{2Q_{L}}}}\cos{\left(\omega\left(T_{k}-T_{j}\right)+\psi\right)}\right).
\end{array}
\label{eq:Vtrans23}
\end{equation}

Eq.~\ref{eq:Vtrans23} now provides us with an explicit relationship between $f_{n,k-1}$ and $f_{n,k}$ that can be used to with the computation of $\langle{I_{th}}\rangle$.

%\begin{widetext}
%\begin{equation}
%\begin{split}
%\textcolor{red}{\textbf{This is Rob's two columns equation}} \\
%E(b_0)=\frac{1}{4} \left(2 (-1)^p (1-2 p) \left(2^{\frac{1}{-2 p-2}} %\left((-1)^p (1-2 p)
%   p\right)^{\frac{1}{-2 p-2}}\right)^{2 p}+2^{\frac{1}{p+1}} \left((-1)^p (1-2 p)
%   p\right)^{\frac{1}{p+1}}\right)
%\end{split}
%\end{equation}
%\end{widetext}

\section{Simulation}

\subsection{ERL BBU code with Filling Pattern}
\begin{figure}
\scalebox{0.5} [0.5]{\includegraphics{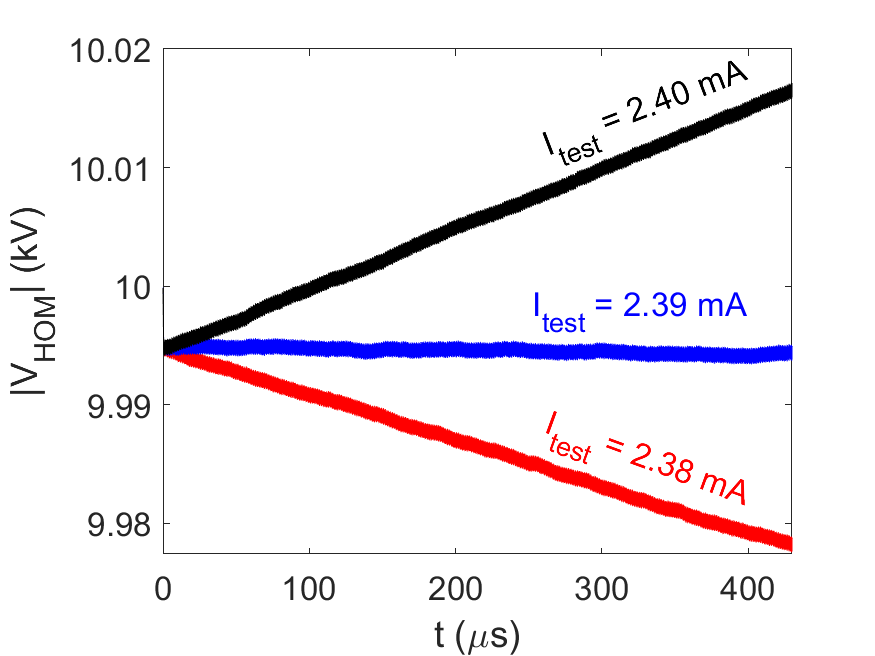}}
	\caption{HOM voltage when test current I$_{test}$ is above (black), below (red), and at (blue) the threshold current.}
	\label{Fig:Benchmark}
\end{figure}

\begin{figure}
\scalebox{0.5} [0.5]{\includegraphics{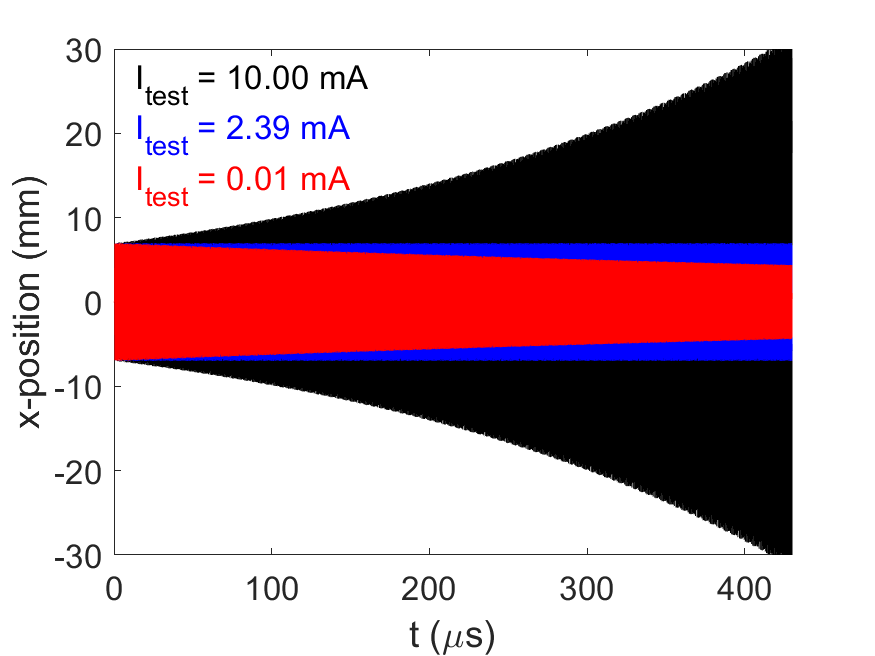}}
	\caption{Particle x-position (in the cavity) when test currents I$_{test}$ is above (black), below (red), and at (blue) the threshold current.}
	\label{Fig:xposition}
\end{figure}

% rephrase so we don't want to give impresson that we have readly available package to download. We have undertaken
As the concept of filling patterns for ERLs is relatively new, none of the existing BBU simulation codes are currently able to incorporate this into their calculations. We have adapted the ERLBBU algorithm~\cite{Pozdeyev2005,TennantDissertation2006,Pozdeyev2006} to include the calculation of BBU with filling patterns by not assuming a constant bunch spacing. These modifications describe the arrival time and energy of each individual bunch. Benchmarking against the experimental results presented in Table 5.1 of Ref.~\cite{TennantDissertation2006} provided consistent results when applying the same bunch pattern assumed in the literature. Our modified script predicted a threshold current of 2.39~mA, which is slightly closer to the experimental results than other codes and analytical estimates due to these corrections. The ERLBBU algorithm starts with a test current and an initial HOM voltage (which for the following simulations is assumed to be 10~kV transverse voltage). Bunches are injected with small, Gaussian distributed, transverse offsets. This simulation estimates if the cavity voltage increases or decreases over time under the test current, as shown in Fig.~\ref{Fig:Benchmark}, and generates a new test current accordingly and repeats the process until the threshold current is determined within a user-defined tolerance range. As seen in Fig.~\ref{Fig:Benchmark}, when the threshold current of 2.39~mA is reached, the cavity HOM voltage has converged to an equilibrium value. The assumed parameters are given in Table~\ref{tab:BBUSimPara}. The magnitude of the transverse offset of the bunches also increases or decreases in a similar fashion to the HOM voltage as shown in the Fig.~\ref{Fig:xposition}.

\begin{table*}
	\caption{6-Turn ERL BBU simulation parameters.}
		\begin{tabular}{lll}
		\hline
			Parameter		~~~&~~~  	{Unit} ~~~&~~~  {Value}        \\
			\hline 
			fundamental mode frequency $f$                  ~~~&~~~ MHz  ~~~&~~~  1497.0  \\
			HOM frequency $f_{HOM}$ 	    ~~~&~~~ MHz ~~~&~~~ 2105.4$-$2106.6 \\
			HOM loaded Q-factor $Q_{L, HOM}$ 	                ~~~&~~~   ~~~&~~~ $6.11\times 10^6$   \\
            HOM geometric shunt impedance $\left(\frac{R}{Q}\right)_{HOM}$     ~~~&~~~ $\Omega$  ~~~&~~~ 29.9   \\
            revolution period for non-transitioning bunches $T_{rev}$   ~~~&~~~ ns  ~~~&~~~  801.67  \\
            revolution period for transitioning bunch $T_{rev, t}$   ~~~&~~~ ns  ~~~&~~~  802.01  \\
%            transfer matrix element $R_{12}$    ~~~&~~~  m      ~~~&~~~   -5.1 \\
            bunch energies at turn 1$-$6        ~~~&~~~  MeV    ~~~&~~~~7.3, 46.3, 85.3, 124.3, 85.3, 46.3   \\
%            injection frequency for single/double spacing $f_{inj}$ ~~~&~~~ $f$  ~~~&~~~  $\frac{1}{5}$/$\frac{1}{10}$  \\
            bunch spacing ~~~&~~~ $T_{RF}$  ~~~&~~~  5 and 10  \\
%            transfer matrix element $R_{12}$    ~~~&~~~  m      ~~~&~~~  -5.1   \\
            injected beam RMS offset $\sigma_{x, offset}/\sigma_{y, offset}$~~~&~~~ $\mu$m  ~~~&~~~  10/1 \\
%            ~~~&~~~         ~~~&~~~      \\
			\hline
		\end{tabular}
	\label{tab:BBUSimPara}
\end{table*}

\subsection{Simulation Initial Parameters}

The scan is performed for a 6-turn ERL with 3 accelerating and 3 decelerating turns. To start with we used similar simulation parameters to those used in Ref~\cite{TennantDissertation2006} as summarized in Table.~\ref{tab:BBUSimPara}. The transitioning bunch is that moving from accelerating to decelerating phase. The transition is achieved by delaying the bunch by least  half an RF cycle. Therefore, the revolution period of the transitioning bunch is half an RF cycle longer. The injected bunches are 7.3~MeV and they gain/lose 39~MeV in each in accelerating/decelerating turn. We simulated two injection frequencies, which are of $\frac{1}{5}^{\text{th}}$ and $\frac{1}{10}^{\text{th}}$ of fundamental mode frequency $f$, i.e. bunches are injected in every 5 and 10 RF cycles $T_{RF}$, respectively. The horizontal injection offsets are set 10 times of the vertical as bunches are bent in the horizontal plane and would be dominated by horizontal jitters. Here we don't need to be the exact on the initial parameters like HOM voltages and offset jitters as we are looking at the trend (growth or decline) to decide whether the test current is above or below $I_{th}$. 

\subsection{Frequency dependence}

\begin{figure}
\scalebox{0.5} [0.5]{\includegraphics{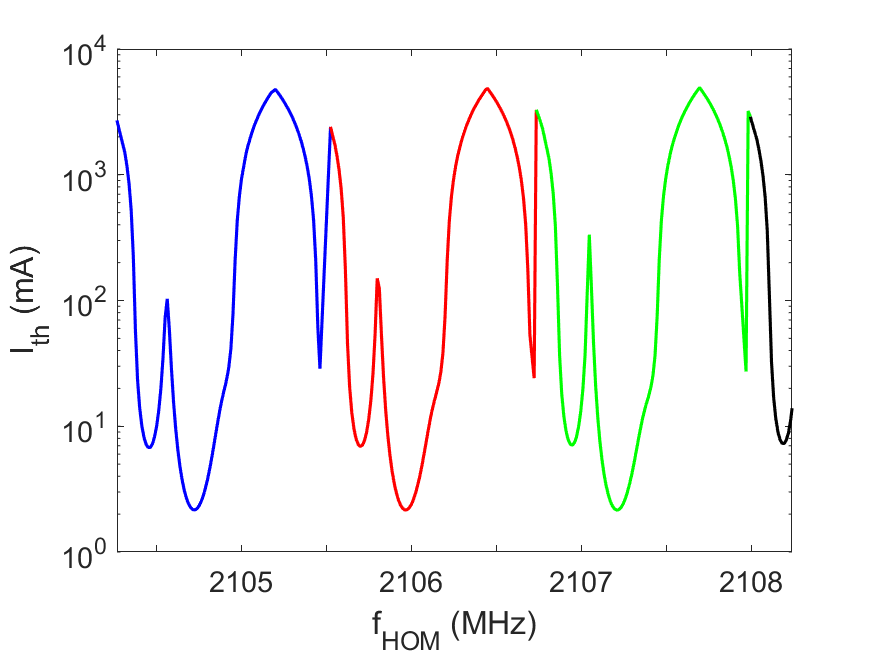}}
	\caption{Threshold current periodicity over HOM frequency. Different period colored differently. The filling pattern used is \{1~2~3~4~5~6\} and the phase advance $\mu = 1.5\pi$.}
	\label{Fig:Ith_Periodicity}
\end{figure}

The results show the threshold current is frequency dependent. Firstly, the simulation results show the threshold current is periodic over HOM frequency range of 1.24~MHz as shown in the Fig.~\ref{Fig:Ith_Periodicity}. The blue, red, and green colors each indicate 1 period of the curve. Such periodicity is expected from the general form the threshold current equation~\cite{TennantDissertation2006}

\begin{equation}
    I_{th} = \frac{-2E_{beam}}{e \frac{\omega_H}{c} \left( \frac{R}{Q} \right)_{HOM} R_{12} sin(\omega_{H} T_{rev})}
\label{eq:IthGen}
\end{equation}

\noindent where $E_{beam}$ is the beam energy in electron Volt, $e$ is electron charge, $c$ is speed of light in vacuum, $\omega_H$ is the HOM angular frequency,  $\left(\frac{R}{Q}\right)_{HOM}$ is geometric shunt impedance of the HOM, $R_{12}$ is the transfer matrix element relates the angular kick to the off-set after one recirculation and $T_{rev}$ is the revolution period. Note that this equation only applicable for estimating threshold current for simple 2-turn ERL. Nevertheless, we can use it to explain periodicity. The period arises from periodicity of the term $sin(\omega_{H} t_{tr})$. Given the revolution period is 802~ns, the period $P_{f,HOM}$ is ${1}/{T_{rev}} = 1.24~\text{MHz}$. Secondly, we see that the periodicity is not exact but the the threshold current slowly decreases as the HOM frequency increases. Such relation can be also seen from the Eq.~\ref{eq:IthGen}. 

%\begin{equation}
%    P_{th,f} = \frac{1}{t_{rev}} = 1.24~\text{MHz}
%\label{eq:IthGen2}
%\end{equation}

\subsection{Filling Pattern Dependence}

We performed ERL BBU tracking simulations and estimated threshold current for different sequence preserving filling patterns, where the cavity will always see same sequence of bunches passing through. The scan is over frequency range of 2.1054 and 2.1066~GHz, which covers one full aforementioned $P_{f,HOM} = 1.24$~MHz period. As the ERL is 6 turn, 6 bunches form a packet. Such packets are repeated multiple times to fill up the circular accelerator ring.

\begin{figure}
	\begin{tabular}{cc}
		\def\stackalignment{l}
		\topinset{\bfseries(a)}{\includegraphics[width=3in]{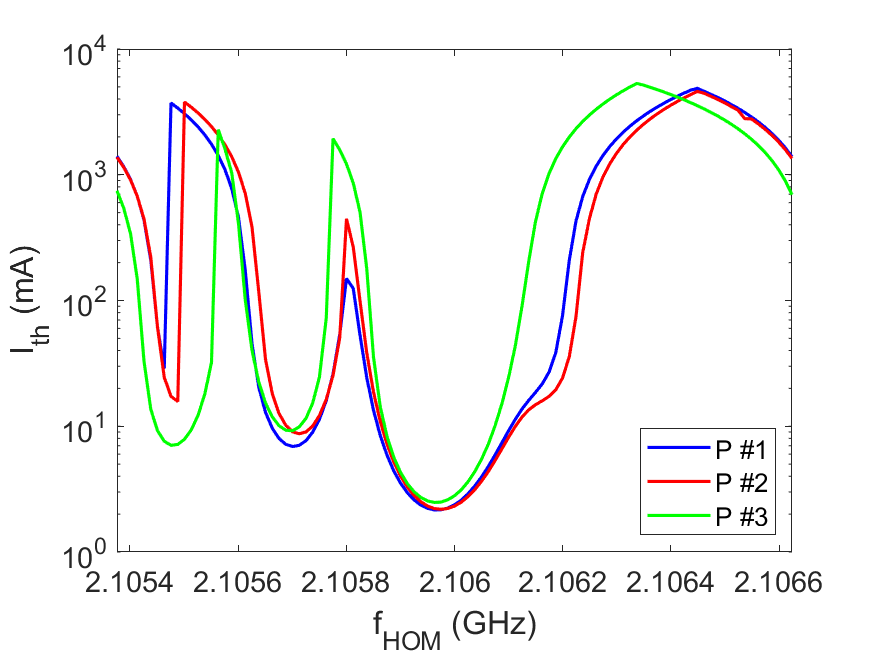}}{0.2in}{0.5in} \\
		\def\stackalignment{l}
		\topinset{\bfseries(b)}{\includegraphics[width=3in]{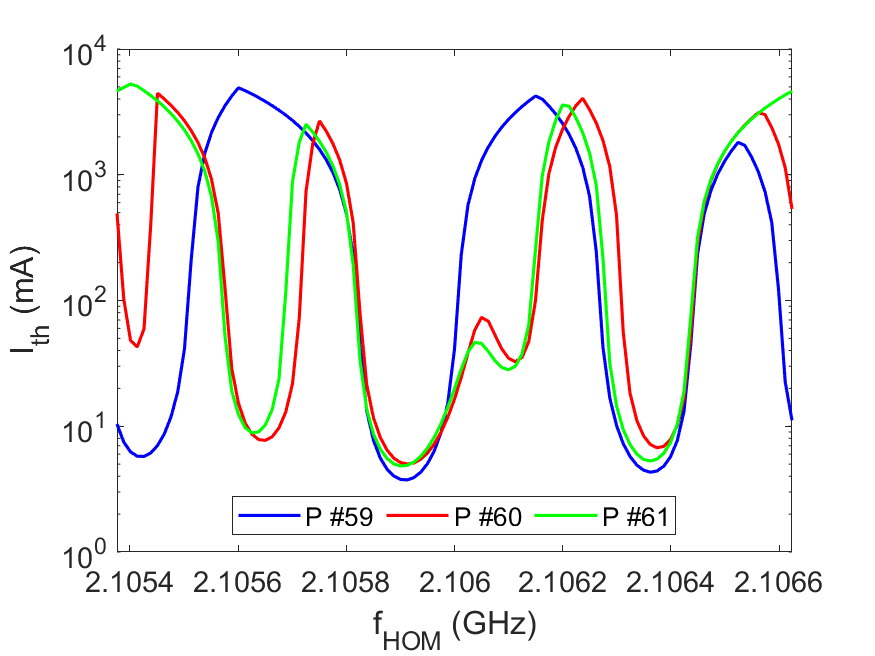}}{0.2in}{0.5in} \\		
	\end{tabular}
	\caption{Filling pattern dependence of the threshold current: (a) filling pattern No. 1$-$3; (b) filling pattern No. 59$-$61. The bunch spacing is $5T_{RF}$.}
	\label{fig:Pat_FHOM_Scan}
\end{figure}

The scan results for $5T_{RF}$ spacing are as shown in the Fig.~\ref{fig:Pat_FHOM_Scan}. The threshold currents as function of HOM frequencies are given for different patterns. As can be seen, each pattern indeed drives BBU instabilities differently and have different threshold currents for same frequency. We do see similar patterns have close $I_{th}$ at some frequencies, as can be seen from the sub-figure (a) for patterns No. 1$-$3. In the case of sub-figure (b), however, we see patterns No. 59 ($\{1~4~3~6~2~5\}$), 60 ($\{1~4~3~6~5~2\}$), and 61 ($\{1~4~5~2~3~6\}$) have mostly different $I_{th}$.

\subsection{Averaged threshold current}

\begin{figure}
\scalebox{0.55} [0.55]{\includegraphics{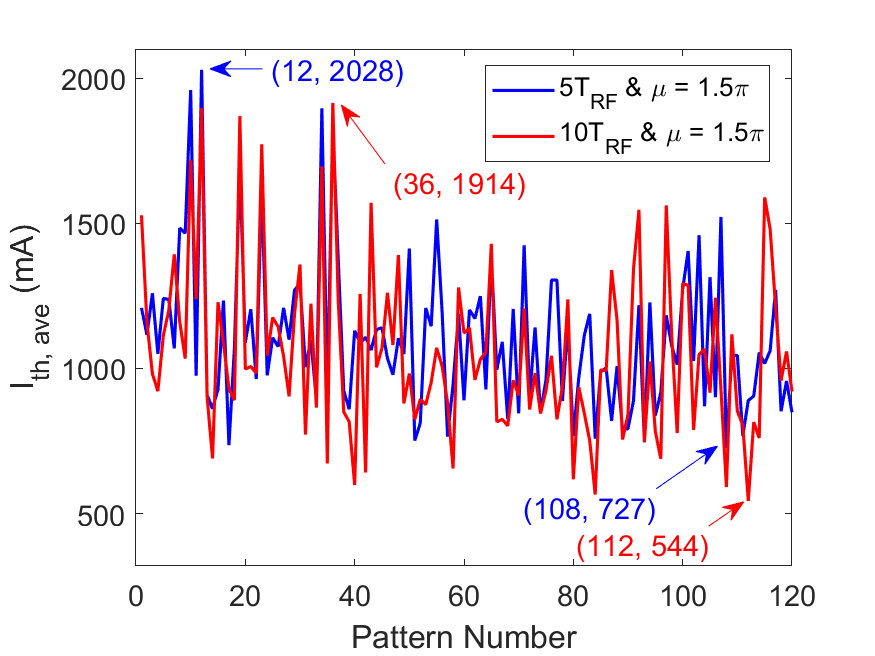}}
	\caption{Frequency averaged threshold current 120 SP patterns. The red and blue curves are the results with $5T_{RF}$ and $10T_{RF}$ bunch spacings, respectively.}
	\label{fig:Ith_vs_Patt} 
\end{figure}

We performed threshold current scans for the 120 SP patterns. The simulated frequency range is 1 period, $P_{f,HOM}$, for each pattern. The averaged threshold current over for each pattern is shown in Fig.~\ref{fig:Ith_vs_Patt} for the case phase advance $\mu$ is 1.5~$\pi$. As can be seen, pattern number 12 has the highest threshold current at 2~A, more than 3 times the lowest threshold current. Pattern numbers 10, 12, 19, 23, 33, and 36 have high threshold currents above 1.6~A for both $10T_{RF}$ and $5T_{RF}$ spacing cases. Clearly, some patterns are better than others in the terms of lowering BBU instabilities. The highest threshold current is 2.028~Amps when filling pattern No. 12 is used. Overall, the $5T_{RF}$ bunch spacings have larger threshold currents as can be seen from Fig.~\ref{fig:Ith_vs_Patt} and Tab~\ref{tab:Iths}. The maximum, minimum, and average of the $5T_{RF}$ bunch spacing are higher than of the $10T_{RF}$.

The relationship between threshold current and filling patterns is complicated as several parameters (like bunch arrival times, energies, and orders) change simultaneously when filling pattern is changed, consequently the average threshold currents appear to be random for different filling patterns. The analytical model fails to predict the threshold current of such a complicated system and one has to rely on numerical simulation tools only.

\begin{table}
	\caption{Maximum, minimum, and average threshold currents for $5T_{RF}$ and $10T_{RF}$ bunch spacings.}
	\begin{ruledtabular}
		\begin{tabular}{lcc}
		 Threshold currents	& Pattern number & Values (mA) \\
			\hline
					$I_{th,~5T_{RF},~max}$ 	& 12  & 2028\\
					$I_{th,~5T_{RF},~min}$	& 108 & 727\\
					$I_{th,~5T_{RF},~ave}$	& --- & 1102\\
					$I_{th,~10T_{RF},~max}$    & 36  & 1914\\
					$I_{th,~10T_{RF},~min}$    & 112 & 544 \\
					$I_{th,~10T_{RF},~ave}$	& --- & 1053 \\
		\end{tabular}
	\end{ruledtabular}
	\label{tab:Iths}
\end{table}

\begin{figure}
	\begin{tabular}{cc}
		\def\stackalignment{l}
		\topinset{\bfseries(a)}{\includegraphics[width=3in]{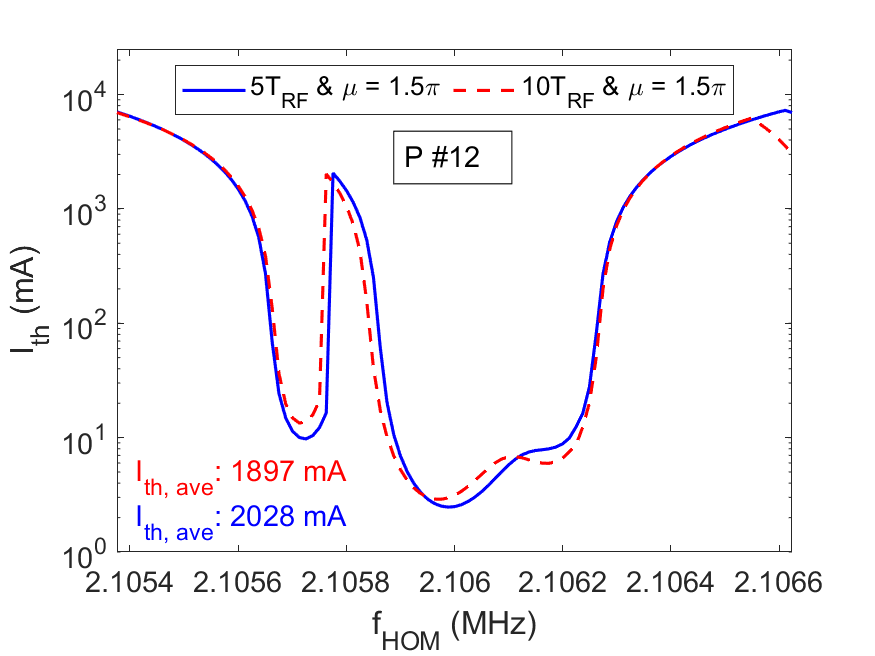}}{0.2in}{2.52in} \\
		\def\stackalignment{l}
		\topinset{\bfseries(b)}{\includegraphics[width=3in]{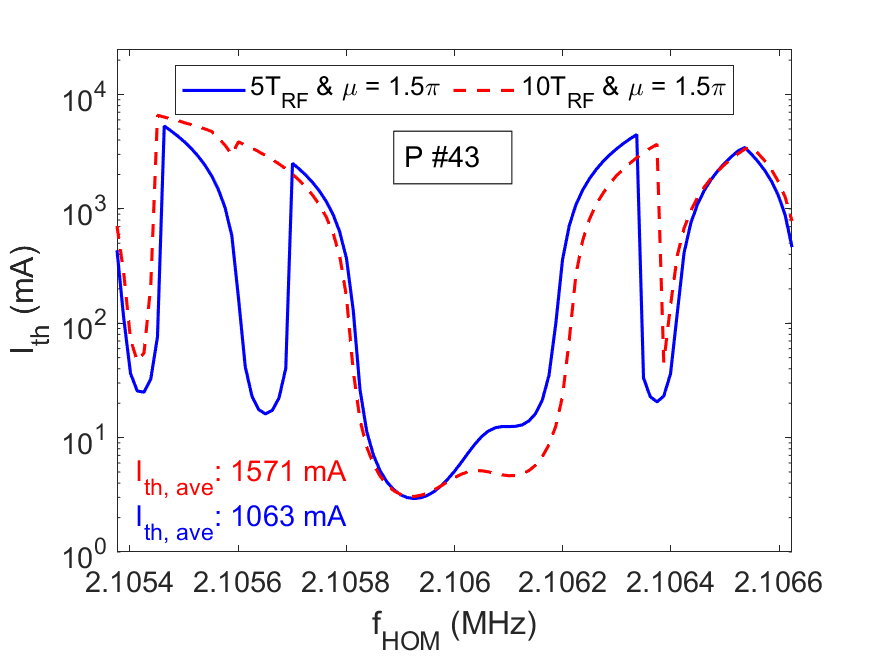}}{0.2in}{2.52in} \\		
	\end{tabular}
	\caption{Threshold currents as function of HOM frequency pattern number 12 and 43 at two bunch spacings. In (a) they similar and in (b) different threshold currents.}
	\label{fig:IthP12P43}
\end{figure}

Another observation is that the spacing of the bunches has a big effect on the threshold currents for some patterns but not others. In pattern number 12 for example, both $5T_{RF}$ and $10T_{RF}$ spacings have similar threshold current, while pattern number 43 for example does not. We also compared frequency scans of these two patterns, as shown in Fig.~\ref{fig:IthP12P43}. 
%For pattern number 12, changing the bunch spacing doesn't have big impact on its threshold current, but it was not the case for the pattern number 43. Again, this shows the relation between threshold current and filling patterns is complicated and bunch spacings play a big role as well.

\subsection{Phase advance dependence of threshold current}
\begin{figure}
\scalebox{0.5} [0.5]{\includegraphics{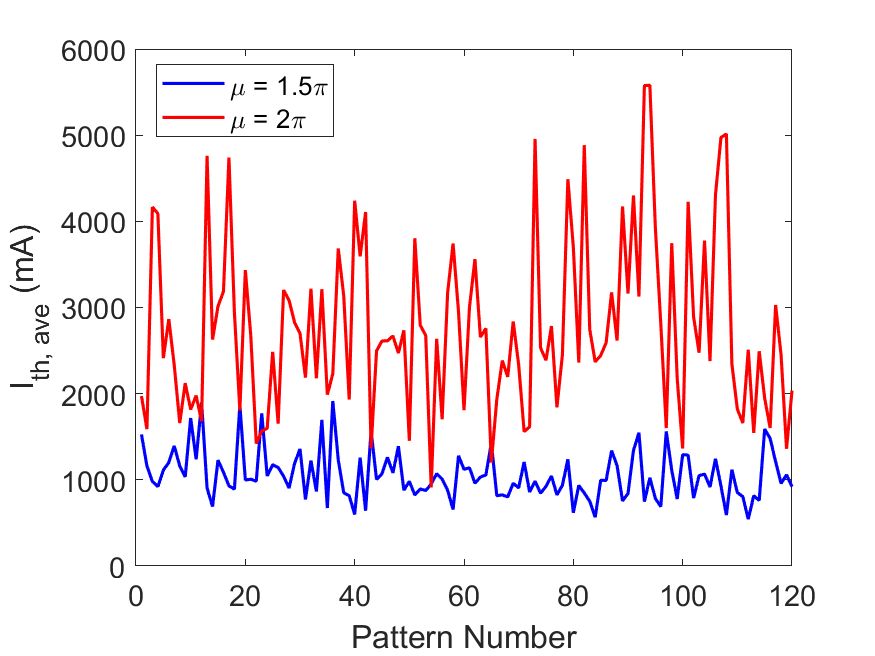}}
	\caption{Average threshold current 120 SP patterns with different phase advances with $10T_{RF}$ bunch spacing. Threshold currents are higher when phase advances are integer times of $\pi$}
	\label{fig:Ith_vs_Patt2} 
\end{figure}

A scan of threshold currents with the two different phase advances of 1.5 and 2~$\pi$ with $10T_{RF}$ bunch spacing is shown in Fig.~\ref{fig:Ith_vs_Patt2}. When the phase advance is integer times $\pi$, bunches enter the cavity with minimum offset and hence the BBU instability is minimized. Consequently, the threshold currents are increased greatly. We see when the tune changed, the dependence of the threshold current on the patterns have changed as well. 

The BBU threshold current is dependent on many parameters that interact in a complicated manner, making it difficult to choose the best pattern with the highest threshold current using the analytical approach. We have however demonstrated that one can find the best filling pattern for a given set of parameters by performing pattern scans with numerical simulations.

\section{Conclusion}
We have investigated the impact of filling pattern choice in a multi-pass ERL on the regenerative BBU instability.
We find analytically that the mean threshold current decreases with increasing ERL turn number, placing constraints on the feasibility of designing arbitrarily many turns. In~\cite{Hoffstaetter2004} states a scaling rule of $I_{th}$ with the number of turns as $1/\left(N\left(2N-1\right)\right)$, which differs slightly from our own form given as $\alpha$ and $\alpha'$ in Eq.~\ref{eq:Vtrans15} and \ref{eq:Vtrans16}, which provides a more generalised expression for this scaling law, although it is also clear that the scaling law will depend on the recirculation scheme. We demonstrated filling pattern dependence of the regenerative BBU instability threshold current. For example, we observed a factor of 3 difference in the threshold current between worst and best patterns for SP schemes. The threshold current and best filling pattern depend on many parameters interacting in a complex manner. We constructed a new ERL tracking code to numerically scan for the best filling pattern to maximise the threshold current. This tool will assist in the design of future ERL projects.

\section{ACKNOWLEDGEMENTS}
The authors would like to thank Dr. Graeme Burt for his useful suggestions and insights. The studies presented have been funded by STFC Grants No. ST/P002056/1 under the Cockcroft Institute Core Grant.

%\section{APPENDIX}
%\bibliographystyle{unsrt}

\end{document}